Communication

# Poly(1-vinyl-1,2,4-triazolium) poly(ionic liquid)s: synthesis and the unique behavior in loading metal ions

Weiyi Zhang, Jiayin Yuan*

Max-Planck-Institute of Colloids and Interfaces, D-14476 Potsdam, Germany
E-mail: Jiayin.Yuan@mpikg.mpg.de

Herein we report the synthesis of a series of poly(4-alkyl-1-vinyl-1,2,4-triazolium) poly(ionic liquid)s either *via* straightforward free radical polymerization of their corresponding ionic liquid monomers, or *via* anion metathesis of the polymer precursors bearing halide as counter anion. The ionic liquid monomers were first prepared *via* *N*-alkylation reaction of commercially available 1-vinyl-1,2,4-triazole with alkyl iodides, followed by anion metathesis with targeted fluorinated anions. The thermal properties and solubilities of these poly(ionic liquid)s have been systematically investigated. Interestingly, it was found that the poly(4-ethyl-1-vinyl-1,2,4-triazolium) poly(ionic liquid) exhibited an improved loading capacity of transition metal ions in comparison with its imidazolium counterpart.

Graphic table of content

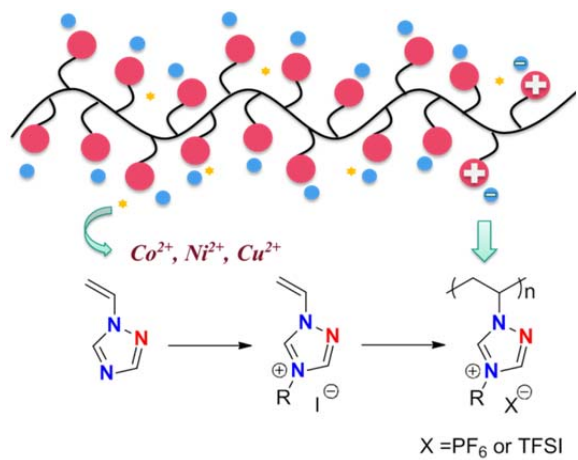

# 1. Introduction

Poly(ionic liquid)s (PILs) are an innovative class of polyelectrolytes prepared *via* polymerization of ionic liquids, or sometimes *via* chemical modification of existent polymers through ion exchange, N-alkylation and ligation of ILs onto functional neutral polymers. In spite of countless structure possibilities up to date, the most common ones originate from the combination of the cations selected from *N,N'*-dialkylimidazolium, *N*-alklypyridinium, tetraalkylammonium, tetraalkylphosphonium, *N,N'*-dialkylpyrrolidinium or 1,2,3-triazolium, with anions chosen from halides ($Cl^-$, $Br^-$, $I^-$), inorganic fluorides ($BF_4^-$, $PF_6^-$), or hydrophobic organics (($CF_3SO_2)_2N^-$, $CF_3SO_3^-$)[1]. Among these examples, the imidazolium-based ones have been studied most systematically as they exhibit the common advantages of ionic liquids and PILs[2]. Nevertheless, the seeking of novel polymer backbones, cations or anions is a continuous effort in the community to enrich the structure toolbox and search for new physical properties and functions. Recently, 1,2,3- and 1,2,4-triazolium-type ionic liquids catch expanding interest due to their exotic structure and potential applications as functional ionic liquids as ionic reaction media and energy rich materials[3]. For instance, Shreeve *et al.* have reported several energetic ionic liquids carrying 1,2,4-triazolium[4]; Rovisere *et al.* utilized 1,2,4-triazolium compounds as N-heterocyclic carbenes's precursor[5]. In the case of triazolium-containing polymers, 1,2,3-triazolium ones are the most dominant. For instance, Drockenmuller *et al.* synthesized a number of PILs derived from 1,2,3-triazolium and discovered them with superior ion conductivity[6]. Polymers containing 1,2,4-triazolium ionic liquid species are rarely reported. To the best of our knowledge, so far only Shreeve *et al.*[7] and Miller *et al.*[8] discussed 1,2,4-triazolium-based polymers and crosslinked PIL networks, respectively. Additionally, an initial study of a 1,2,4-triazolium-3-thiolate compound was reported by Altland *et al.* without discussion of its polymerization.[9] It is in our opinion a group of PILs (or polymers in general) to be explored, where new knowledge and properties are expected to emerge. Herein we report the synthesis of a series of 1-vinyl-1,2,4-triazolium-based PILs bearing bis(trifluoromethanesulfonyl)imide ($TFSI^-$) and hexafluorophosphate ($PF_6^-$) as counter anions either through straightforward radical polymerization of the corresponding ionic liquid monomers or through anion metathesis of the polymer precursors bearing iodide anion. Compared with analogues of imidazoliums, poly(1-vinyl-1,2,4-triazolium)s show unique arrangement of nitrogen atoms in the cation ring structure, which exhibits, as proven here, enhanced coordination strength with transition metal ions.

# 2. Experimental Section

Materials, synthetic procedure, analytical instrumentation and additional experimental data are provided in supporting information.

## 3. Results and discussion

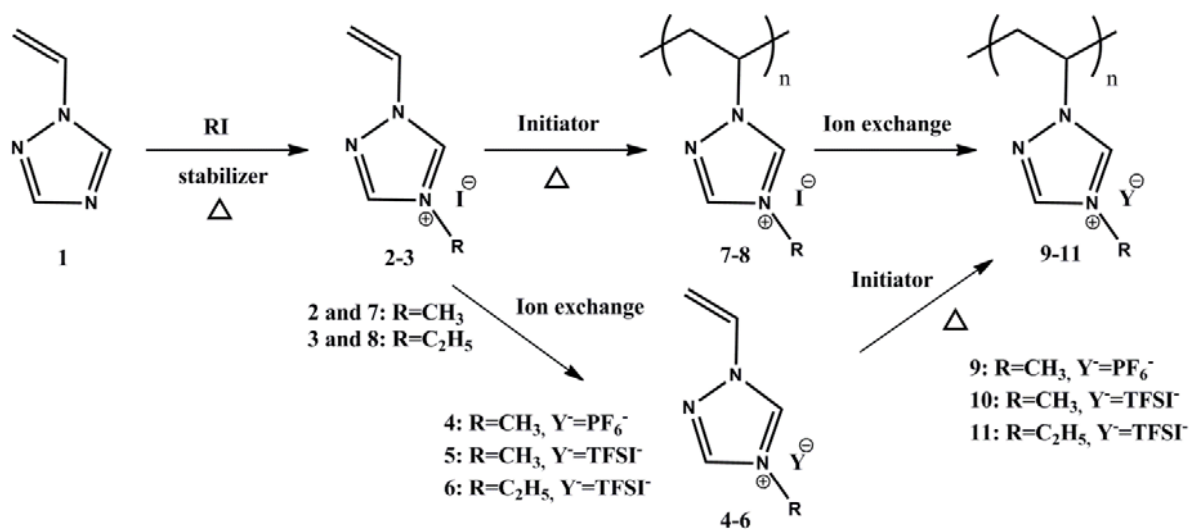

**Scheme 1.** Synthetic route towards poly(4-alkyl-1-vinyl-1,2,4-triazolium) PILs.

Poly(4-alkyl-1-vinyl-1,2,4-triazolium iodide)s with methyl (**7**) and ethyl (**8**) substituents were first synthesized in two steps (**Scheme 1**). Firstly, iodide monomers **2** and **3** were achieved by *N*-alkylation of 1-vinyl-1,2,4-triazole with alkyl iodides. The preference of *N*-alkylation reaction specifically at the 4-position nitrogen in the 1,2,4-triazolium ring has been proven by previous literatures, for example the resonance structure effect reported by Texter[10], further on we confirmed the structure of monomer **2** quaternized at the 4-position nitrogen by its proton-carbon heteronuclear multiple bond correlation ($^1$H-$^{13}$C HMBC) (**Figure S5**) and single crystal structure data. As can be seen in **Table S1**, monomer **2** bearing a methyl substitute is a monoclinic crystal with cell parameters of $a$=5.0409(5) Å, $b$=25.591(4) Å, $c$=6.9793(8) Å and $\beta$=103.817(11)°. Accordingly, this single crystal structure (CCDC 1436278) can be assigned to the P2(1)/c group. **Figure 1** displays the packing structures of monomer **2** constructed from views of three individual axes. The preferential *N*-alkylation at 4 position nitrogen in our opinion is a nature outcome of multiple factors, *i.e.* a higher electron density and a less sterical hindrance at 4 other than 2 position nitrogen, as well as a favorable resonance structure in the *N*-alkylation product at 4 position. After solving the concern of the *N*-alkylation position with alkyl iodide, the chemical structure of monomers **2** and **3** were further confirmed by $^1$H NMR and $^{13}$C NMR spectroscopy, respectively (Supporting Information). In **Figure 2**, the $^1$H NMR spectrum of 1-vinyl-1,2,4-triazole (**A**) was used as a reference, to which its methylation product, monomer **2** (**B**) was compared. 1-Vinyl-1,2,4-triazole (**A**) is equipped with five protons from 3 vinyl ones (c, d, e at 7.33, 5.71 and 5.00 ppm, respectively) and 2 triazole ones (a at 8.78 and b at 8.10 ppm), which after methylation with methyl iodide all shifted to low field. In monomer **2** (**B**) additionally a new peak f' came up and can be allocated to the protons in the methyl group adjacent to the *N*-alkylated nitrogen. The integration ratio of f'/e' (3.0) matches the theoretic value of a full *N*-alkylation, and verified the quantitative methylation. By anion exchange of monomer **2** and **3** with salts, 1,2,4-triazolium monomers with fluorinated anions of $PF_6^-$ or $TFSI^-$ (**4-6**) were synthesized, the $^1$H NMR spectra of which (data not shown) are similar to their halide monomers.

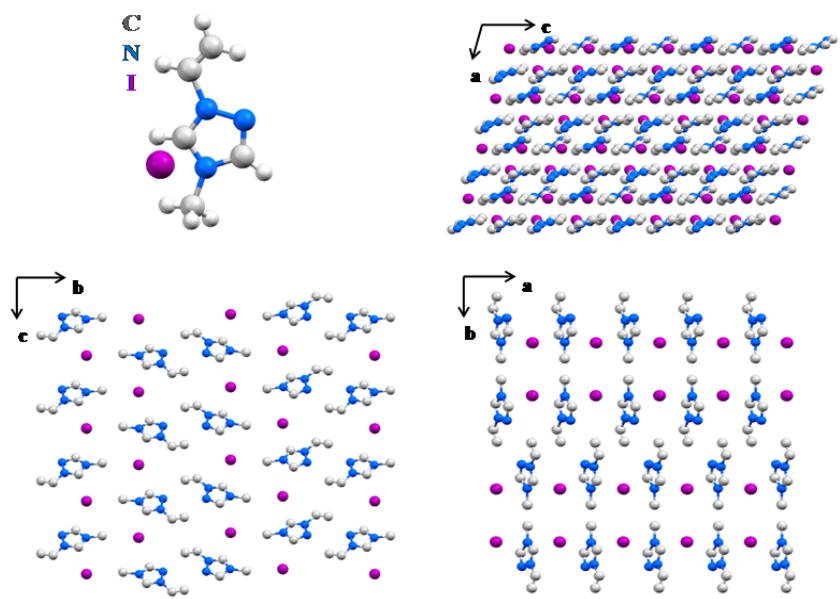

**Figure 1.** Single crystal structure and crystallographic packing structure of monomer **2** (CCDC 1436278 contains the supplementary crystallographic data for this article. These data can be found from The Cambridge Crystallographic Data Centre via www.ccdc.cam.ac.uk/data_request/cif.)

Free radical polymerization of monomers **2** and **3** leads to the formation of polymers **7** and **8**, which were purified *via* dialysis to remove residue monomers. The successful polymerization was confirmed by $^1$H NMR spectroscopy (Supporting Information). Here, $^1$H NMR spectrum of polymer **7** was illustrated in Figure 2 (**C**). It is clearly seen that proton peaks of vinyl group (c', d', e' at 7.55, 6.02 and 5.58 ppm, respectively) vanished after polymerization as expected, and new proton peaks (c" at 4.60 and e" at 2.34 ppm ) appeared, which are assigned to the backbone protons of the polymer. Gel permeation chromatography (GPC) measurement offers a further proof of polymeric natures for both polymer **7** and **8**. The measurement was conducted in an aqueous solution of methanol using acetic acid as buffer. Both polymers displayed in the GPC trace a monomodal distribution and number-averaged apparent molecular weight of 47 and 55 kg/mol, respectively (**Table S2**). Similarly, polymers **9-11** bearing $PF_6^-$ or $TFSI^-$ anion were prepared *via* radical polymerization from their monomers **4-6**. Worth mentioning is that these three polymers can also be accessed *via* post-polymerization anion exchange of polymer **7** or **8**. As previously reported, synthesis of ionic polymers *via* anion exchange may not ignore the residue halide contaminant unfavorable in some applications.

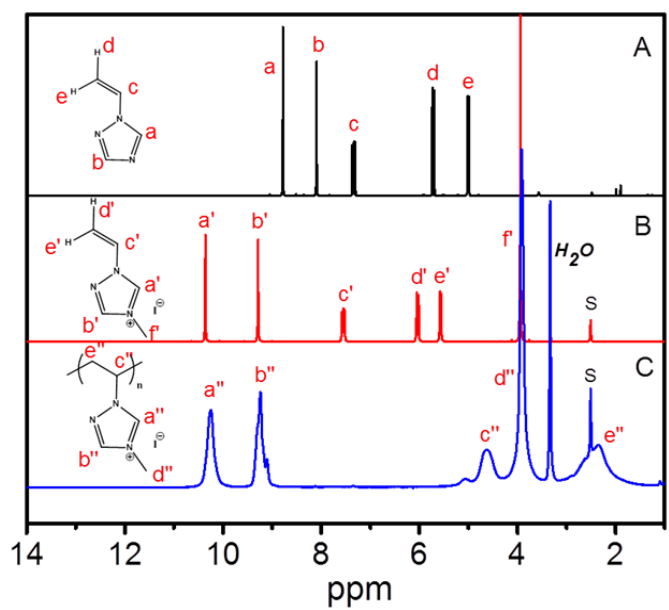

**Figure 2.** $^1$H-NMR spectra of (A) 1-vinyl-1,2,4-triazole; (B) 4-methyl-1-vinyl-1,2,4-triazolium iodide (monomer **2**); and (C) poly(4-methyl-1-vinyl-1,2,4-triazolium iodide) (polymer **7**).

Normally, the thermal properties of ionic liquids and their polymers strongly depend on the interactions between cations and anions. By changing the types of cations or anions the melting points, thermal stability and solubility can be tuned. As detected from differential scanning calorimetry (DSC, **Figure 3**), monomer **2** melts at 114.5 °C and remains stable up to 140 °C before the appearance of a large exothermal peak resulting from the thermal polymerization; notably when replacing the methyl by ethyl substituent, monomer **3** reaches a higher melting point at 145 °C. Comparably, monomer **4** bearing $PF_6^-$ anion melts at a much lower temperature of 72 °C. Monomer **5** and **6** display rather low glass transition temperature ($T_g$) at -61 and -67 °C, respectively, thus being a viscous liquid at room temperature. In the strict sense of the definition of ionic liquids, only monomer **4**, **5** and **6** are considered as ionic liquids, since they reach a liquid state below 100 °C. Correspondingly their polymers **9-11** are classified as PILs, which are studied in detail later. In **Figure S3**, the DSC curves of all polymers are presented. Polymers **10** and **11** carrying TFSI⁻ anion show the $T_g$ at 90 and 60 °C, respectively, while $T_g$ in polymer **9** bearing PF$_6^-$ anion and the iodide-containing polymers **7** and **8**, is absent up to 200°C. This trend is expected, as the large sized TFSI⁻ anion is well-known due to its plasticizing effect to increase the free volume and thus shift $T_g$ to relatively low temperature that can be detected by DSC. The thermal stability of PILs (**9-11**) as well as their precursors (**7** and **8**) was determined by thermogravimetric analysis (TGA, **Figure S2**) performed under N$_2$ flow. The decomposition (defined as 5 wt% mass loss) of iodide-containing polymers **7** and **8** are 237 and 239 °C, respectively, generally lower than the PILs bearing fluorinated anions, 266 °C for **9**, 313 °C for **10** and 283 °C for **11**. This result is expected, as iodide is a stronger nucleophile than PF$_6^-$ and TFSI⁻, which attacks the imidazolium ring at elevated temperature[11].

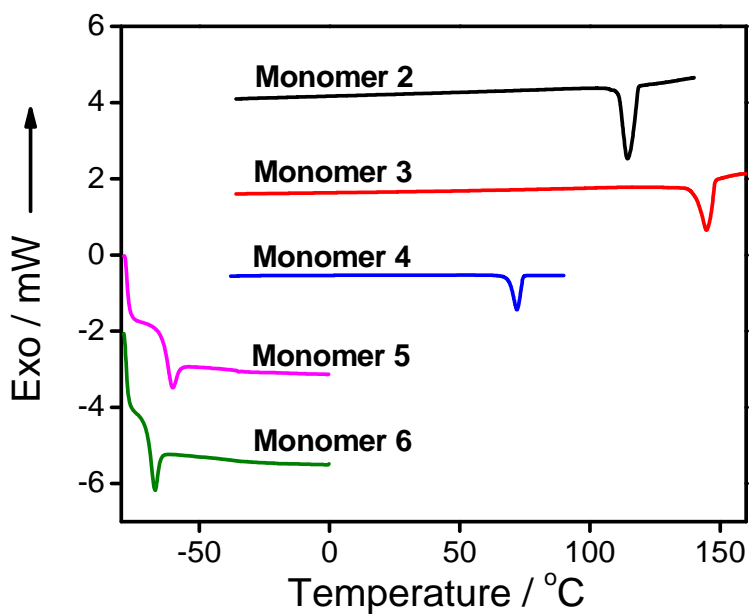

**Figure 3.** DSC curves of monomers **2** (black), **3** (red), **4** (blue), **5** (pink) and **6** (green). Heating rate: 10 K/min under nitrogen.

As the successful anion metathesis for both monomers (**2** and **3**) and polymers (**7** and **8**) cannot be verified by $^1$H NMR, ATR/FTIR spectroscopy measurements were conducted to support the occurrence of the anion exchange reaction (**Figure S4**). Since iodide anions are invisible in ATR/FTIR, only the absorption bands of poly(4-alkyl-1-vinyl-1,2,4-triazolium) cation can be recorded in the spectra. In the case of monomers and polymers carrying TFSI⁻ as anion, their IR spectra have been systematically investigated in previous literatures[12]. On account of that, for monomer **4** and polymer **8**, three peaks at 1177, 740, and 569 cm$^{-1}$ can be clearly assigned to $v_a$CF$_3$, $\delta_s$CF$_3$, and $\delta_a$CF$_3$ of the TFSI anion, respectively. Also four bands of stretching ($v_a$SO$_2$ at 1343 cm$^{-1}$, $v_s$SO$_2$ at 1133 cm$^{-1}$) and bending ($\delta_a$SO$_2$ at 611 cm$^{-1}$ and shoulder $\delta_s SO_2$ at 600 cm$^{-1}$) of =SO$_2$ groups of TFSI$^-$ can be recognized. Furthermore, asymmetric (1051 cm$^{-1}$), symmetric (764 cm$^{-1}$) stretching bands of S-N-S group, and symmetric stretching of C-S (790 cm$^{-1}$) were spotted as expected, confirming the TFSI$^-$ anion as well. For monomer **4** and polymer **9** carrying PF$_6^-$ as anion, the broad peak at 820 cm$^{-1}$ can be assigned to the symmetric stretching of the PF$_6^-$ anion[13]. Finally, the disappearance of stretching bands (vinyl C-H at 3100 cm$^{-1}$) showed difference between the monomers and the polymers. Therefore, ATR/FTIR spectra offered solid evidence of the anion exchange. It should be mentioned that the ATR/FTIR analysis here is only qualitative. To determine the purity of the monomers and polymers after anion exchange, titration was performed via using a silver nitrate reagent, which proves a quantitative anion exchange in the monomers (> 95%) and a high efficiency of anion exchange in the polymers (> 92%, **Table S3**).

Altering the counter anions and alkyl chains of ionic liquids and PILs can impact their solubility properties. **Table 1** summarizes the solubility of 1,2,4-triazolium-based monomers and polymers tested under ambient temperature at a concentration of 1.0 wt%. Independent of the anion exchange, all monomers and polymers were found insoluble in chloroform and toluene, and soluble in DMF and DMSO. For other solvents tested, in general, when polymers or monomers bear iodide and short alkyl substitutes, they are normally soluble in water and polar organic solvents. Upon exchanging iodide to larger, fluorinated and hydrophobic ones, the products turned soluble only in organic solvents. For monomer **2**, it can be dissolved in water, methanol and acetone. In the case of monomer **3**, as the alkyl length extends from methyl to ethyl, *i.e.* increasing slightly its hydrophobicity, it becomes insoluble in water and gradually dissolvable in tetrahydrofuran (THF), chloroform (CHCl$_3$) and ethyl acetate (EtOAc). When fluorinated, hydrophobic anions (PF$_6^-$ or TFSI$^-$) were incorporated, the monomers can be dissolved in all polar organic solvents; unexpectedly, monomer **4** carrying PF$_6^-$ anion maintains its solubility even in water, as PF$_6^-$ is less hydrophobic than TFSI$^-$. Polymerization of monomers can alter their solubility. For example, iodide-containing polymers **7** and **8** become only soluble in water and DMF/DMSO; polymer **9** (bearing PF$_6^-$) even cannot be dissolved in any solvents other than DMF/DMSO. Interestingly, polymer **10** and **11** (bearing TFSI$^-$) maintained the same solubility as their monomers.

Polymer **11** (bearing TFSI$^-$) was further compared with poly(1-vinyl-3-ethylimidazolium TFSI) (PEVImTFSI), whose chemical structure is of high similarity to polymer **11**, except the replacement of one carbon atom in the imidazolium ring by nitrogen. We found that polymer **11** exhibits a general similar solubility to PEVImTFSI expect in methanol, which dissolves polymer **11** but not PEVImTFSI[14]. Indeed the single atom change in the 5-member cation ring from imidazolium to 1,2,4-traizolium does influence the physical properties.

*Table 1.* Solubility of 1,2,4-triazolium monomers and polymers in different solvents.

| Compound | | H$_2$O | MeOH/acetone/ACN | THF | CHCl$_3$/toluene | EtOAc | DMF/DMSO |
|---|---|---|---|---|---|---|---|
| **Monomer** | 2 | + | + | - | - | - | + |
| | 3 | + | + | - | - | - | + |
| | 4 | + | + | + | - | + | + |
| | 5 | - | + | + | - | + | + |
| | 6 | - | + | + | - | - | + |
| **Polymer** | 7 | + | - | - | - | - | + |
| | 8 | + | - | - | - | - | + |
| | 9 | - | - | - | - | - | + |
| | 10 | - | + | + | - | + | + |
| | 11 | - | + | + | - | - | + |

*(+) – soluble at 1.0 wt%; (-) - insoluble at 1.0 wt%*

PILs are capable of binding transition metal ions *via* coordination, which is of significant value in numerous practical applications, such as catalysis, selective sorption, metal batteries, just to name a few. A high capacity of loading metal ions by PILs is important for some applications. For instance, imidazolium-based porous PILs that was able to load as much as 25 wt% of CuCl$_2$, presented an excellent efficiency in catalyzing aerobic oxidation of hydrocarbons[15]. This finding inspired us to investigate the adsorptive capability of poly(1-vinyl-1,2,4-triazolium) PILs towards transition metal ions because of the extra lone electron pair in 2-position nitrogen for coordination with metal species. Here, PEVImTFSI as one of the most popularly studied PILs, was used as reference to compare with polymer **11** due to their structural similarity. Three metal salts, *i.e.* Co(TFSI)$_2$, Ni(TFSI)$_2$, and Cu(TFSI)$_2$, which feature the same anion TFSI as polymer **11** to skip the anion factor, were tested. Experimentally polymer **11** and PEVImTFSI were mixed individually with Co(TFSI)$_2$, Ni(TFSI)$_2$, or Cu(TFSI)$_2$ in DMF. The molar ratio of (Co, Ni, or Cu)/(imidazolium or 1,2,4-triazolium) is set to 2. After stirring overnight, samples were exhaustively dialyzed to remove unloaded metal salts. The final polymer/metal complex products were dried under high vacuum (1x10$^{-3}$ mbar) at 100 $^{\circ}$C for 24h. The residue solid was examined by elemental analysis to determine the metal content, which is shown in **Figure 4**. Two obvious trends are visible: first, independent on the type of metal ions, polymer **11** presents a higher loading capacity than PEVImTFSI, here specifically 10%, 50% and 20% more in the case of Co(II), Ni(II) and Cu(II), respectively. As reported previously, 1,2,4-triazolium ions normally possess stronger acidity than their imidazolium analogues,[16] and deprotonate easier in DMF, a weak or neutral Lewis base, thus coordinating stronger with transition metal ions. Nevertheless, no experimental proof has been provided to this polymeric system so far, which remains a future task. In addition, the higher loading capacity of Co(II) than Cu (II) and Ni(II) in both polymer **11** and PEVImTFSI might stem from their coordination structures, in which Co(II) is known to build up an octahedral ligand field, while Cu(II) and Ni(II) prefer square planar or tetrahedron configuration. It is therefore concluded that poly(1-vinyl-1,2,4-triazolium) PILs undergo stronger coordination with transition metal ions than polyvinylimidazolium PILs. Second, Co(II) ions complex much stronger than Ni(II) and Cu(II), independent of the type of the PILs tested here. Thus, the metal loading in PILs is ion-specific.

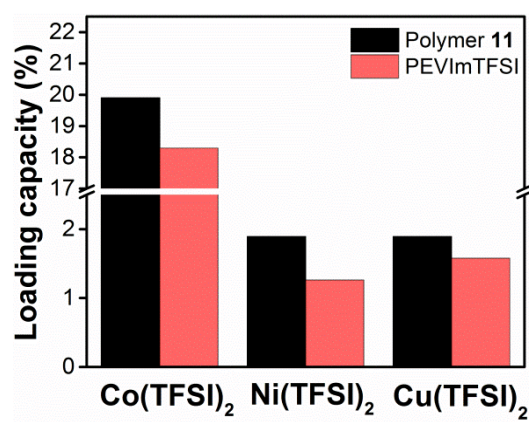

**Figure 4.** Loading of Co/Ni/Cu ions in polymer **11** (black) and PEVImTFSI (red) at a fixed mixing molar ratio of metal/monomer unit ~ 2. The y axis, loading capacity denotes the molar percentage of the loaded metal ions with regard to the monomer unit.

## 4. Conclusions

In summary, we reported a series of 4-alkyl-1-vinyl-1,2,4-triazolium-based ionic liquids and poly(ionic liquid)s. A systematic exploration of their physicochemical properties has been conducted. Similar to common ionic liquids, the melting points of 1,2,4-triazolium monomers can be tailored by selecting different anions. What is very interesting is that the 1,2,4-triazolium PILs are different from their imidazolium analogues in immobilizing transition metal ions, which makes this group of PILs appealing in, *e.g.* catalysis and sorption. Moreover, the high content of nitrogen in the 1,2,4-triazolium ring may be suitable candidates for nitrogen-doped carbons.

## Supporting Information

Supporting Information is available from the Wiley Online Library or from the authors.

Acknowledgements: The authors acknowledge financial support from the Max Planck Society and the grant of China Scholarship Council. We thank Prof. Feihe Huang for the single crystal structure measurements.

**Keywords**: poly(ionic liquid); poly(1-vinyl-1,2,4-triazolium); metal adsorption

# Poly(1-vinyl-1,2,4-triazolium) poly(ionic liquid)s: synthesis and the unique behavior in loading metal ions

Weiyi Zhang, Jiayin Yuan*

## Experimental Section

### Materials

1-Vinyl-1,2,4-triazole (**1**, 98%), iodomethane ($CH_3I$, 99%), iodoethane ($CH_3CH_2I$, 99%), butylated hydroxytoluene (BHT, 98%), N,N'-azobisisobutyronitrile (AIBN, 98%, recrystallized from methanol), potassium hexafluorophosphate ($KPF_6$, 98%), and all organic solvents were purchased from Sigma-Aldrich. 2,2'-Azobis[2-methyl-N-(2-hydroxyethyl) propionamide] (VA086, 98%) was purchased from Wacker-Chemie. Chloroethane ($CH_3CH_2Cl$, ca. 15% in tetrahydrofuran; 2 mol/L), and lithium bis(trifluoromethylsulfonyl)imide (LiTFSI, >98%) were obtained from TCI chemicals.

### Characterization Methods

Thermogravimetric analysis (TGA) experiments were measured and recorded on a Netzsch TG209-F1apparatus under a heating rate of 10 K $min^{-1}$ with constant nitrogen flow. Differential scanning calorimetry (DSC) measurements were carried out under nitrogen flow on a Perkin-Elmer DSC-1 instrument at a heating/cooling rate of 10 K $min^{-1}$. The melting points were determined from the heating curves and they were measured in the second heating cycle. Fourier transform infrared spectroscopy (FT-IR) spectra were accomplished on a BioRad 6000 FT-IR spectrometer; samples were measured in solid states using a Single Reflection Diamond ATR. $^1$H NMR and $^{13}$C NMR spectra were performed on a Bruker DPX-400 spectrometer in deuterated solvents using residual solvent as reference. Gel permeation chromatography (GPC) was performed using NOVEMA Max linear XL column with a mixture of 80% of aqueous acetate buffer and 20% of methanol. Conditions: flow rate 1.00 mL $min^{-1}$, PSS standards using RI detector-Optilab-DSP-Interferometric Refractometer (Wyatt-Technology). Metal elemental analyses were measured by company Mikroanalytisches Laboratorium Kolbe (www.mikro-lab.de).

**Synthesis of 4-methyl-1-vinyl-1,2,4-triazolium iodide (2)**

A 100 mL flask was filled with a mixture of 1-vinyl-1,2,4-triazole **1** (5mL, 5.5g, 57.83 mmol), iodomethane (5.4 mL, 12.3 g, 86.74 mmol) and 2,6-di-tert-butyl-4-methylphenol (50 mg, 0.227 mol). After stirring and heating for 24 h at 50 °C, the crude product was filtered off and washed with THF for three times. A pale yellow solid (13.21 g, 96.4%) was obtained. $^1$H NMR (400 MHz, DMSO-$d_6$, δ, ppm): 10.36 (s, 1H), 9.28 (s, 1H), 7.55 (dd, 1H, $J_1$=16 Hz, $J_2$=8 Hz), 6.03 (d, 1H, $J$=16 Hz), 5.57 (d, 1H, $J$=8 Hz), 3.94(s, 3H); $^{13}$C NMR (400 MHz, DMSO-$d_6$, δ, ppm): 146.02, 142.67, 129.63, 110.85, 35.34.

**Synthesis of 4-ethyl-1-vinyl-1,2,4-triazolium iodide (3)**

Following the same procedures described in the synthesis route of monomer **2**, a pale yellow solid (13.76 g, 94.8%) was obtained. $^1$H NMR (400 MHz, DMSO-$d_6$, δ, ppm): 10.57 (s, 1H), 9.44 (s, 1H), 7.46 (dd, 1H, $J_1$=16 Hz, $J_2$=8 Hz), 6.01 (d, 1H, $J$=16 Hz), 5.57 (d, 1H, $J$=8 Hz), 4.32 (q, 2H, $J$=8 Hz), 1.49 (t, 3H, $J$=8 Hz); $^{13}$C NMR (400 MHz, DMSO-$d_6$, δ, ppm): 145.04, 141.94, 129.66, 110.97, 44.15, 14.70.

**Synthesis of poly(4-methyl-1-vinyl-1,2,4-triazolium iodide) (7)**

A mixture of 5 g (21.10 mmol) of monomer **2**, 50 mg (0.3 mmol) of AIBN, and 20 mL of anhydrous DMF was put inside a 50 mL schlenk flask under argon protection. Three freeze-pump-thaw cycles were applied for oxygen removal. The reaction was stirred at 75 °C for 24 h and the crude product was dialyzed in water for two days. A yellow powder was received after removal of solvents, the yield was 3.6 g (72%). $^1$H NMR (400 MHz, DMSO-$d_6$, δ, ppm): 10.25 (br, 1H), 9.24 (m, 1H), 4.62 (br, 1H), 3.91 (br, 3H), 2.34 (br, 2H).

**Synthesis of poly(4-ethyl-1-vinyl-1,2,4-triazolium iodide) (8)**

Following the same procedures described in synthesis of polymer **7**, a yellow powder was obtained with a yield of 3.25 g (65%). $^1$H NMR (400 MHz, DMSO-$d_6$, δ, ppm): 10.51 (br, 1H), 9.33 (m, 1H), 4.95 (m, 1H), 4.31 (br, 2H), 2.55 (br, 2H), 1.56 (br, 3H).

**Synthesis of 4, 5, and 6 *via* anion exchange**

Anion exchange was performed by dropwise addition of solution 2) into solution 1) : 1) 1 g of monomer **2** or **3** dissolved in 50 mL of deionized water, and 2) 1.05 eq. of potassium hexafluorophosphate or lithium bis(trifluoromethanesulfonyl)imide in 20 mL deionized water.

For monomer **4**, after vigorous stirring, it was extracted by ethyl acetate, followed by evaporation to remove the solvent to yield pale yellow product. For monomer **5** or **6**, white precipitate appeared, and was filtered off and washed with deionized water for 3 times, before it was dried at 40°C under high vacuum.

## Synthesis of 9, 10, and 11 *via* anion exchange

Anion exchange was performed by dropwise addition of solution 2) into solution 1) : 1) 1 g of polymer **7** or **8** dissolved in 50 mL of deionized water, and 2) 1.05 eq. of potassium hexafluorophosphate or lithium bis(trifluoromethanesulfonyl)imide in 20 mL deionized water. White precipitate appeared, and was filtered off and washed with deionized water for 3 times, before it was dried at 100 °C under high vacuum.

## Solubility tests

10 mg of the monomer or polymer was added to 1 mL of solvent (water, methanol, acetone, THF, ethyl acetate, toluene, chloroform or DMF). The sample was shaken fiercely for 10 min before the solubility at ambient temperature was checked.

## Analysis of $^1$H-$^{13}$C HMBC spectrum for monomer 2

In order to clarify the *N*-alkylation position, the $^1$H-$^{13}$C heteronuclear multiple-bond correlation (HMBC) spectrum of monomer 2 has been conducted. In **Figure S5**, carbon 1 and 2 are coupled with proton 2 and 1, respectively; and carbon 1 is also coupled with proton 3, which illustrates carbon 1 can only be the one closer to the vinyl group, since other carbons do not have the possibility to couple with proton 3 (except carbon 4). For carbon 3, it is coupled both with proton 1 and proton 4 and 5, which defines its only possible position of the vinyl group as the carbon directly attached to nitrogen; consequently carbon 4 can be assigned to the rest carbon of the vinyl group, as it is coupled only with proton 3. Now carbon 5 can only be assigned to the last carbon as the methyl adduct. If carbon 5 was attached to N2, only carbon 2 would be coupled with proton 6, which is not observed, thus N2 alkylation is excluded; when carbon 5 is the one connected with N4, proton 6 will be coupled with both carbon 1&2, which is indeed observed. As a result, the spectrum is fitted with N4 alkylation.

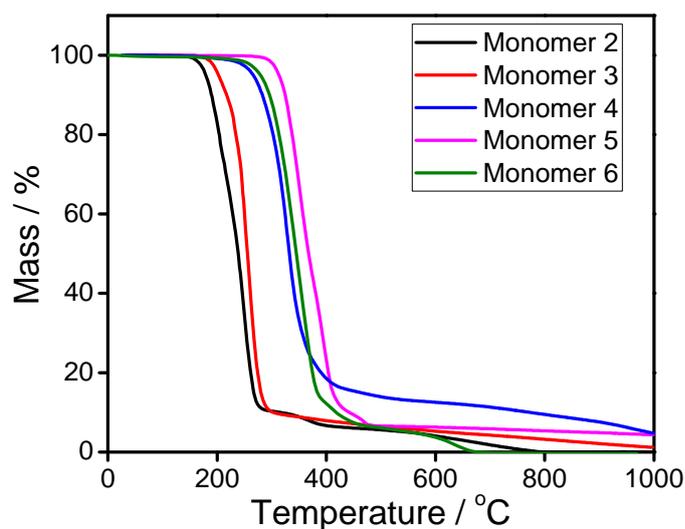

**Figure S1.** TGA curves of 4-alkyl-1-vinyl-1,2,4-triazolium monomers. Heating rate: 10 K/min under nitrogen.

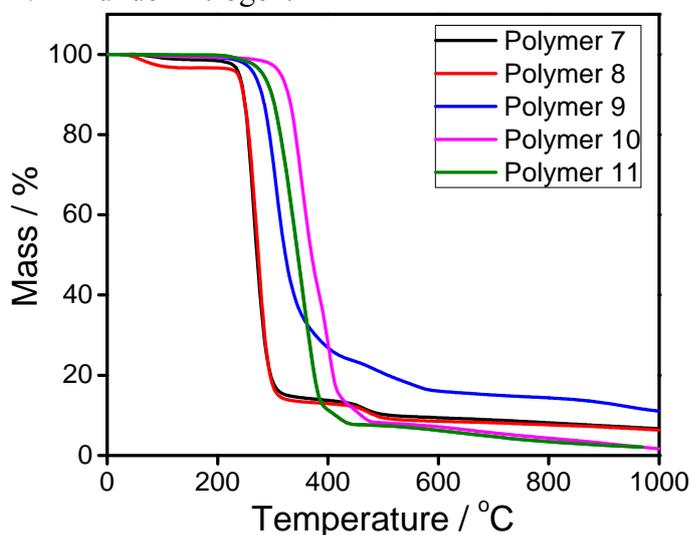

**Figure S2.** TGA curves of poly(4-alkyl-1-vinyl-1,2,4-triazolium)s. Heating rate: 10 K/min under nitrogen.

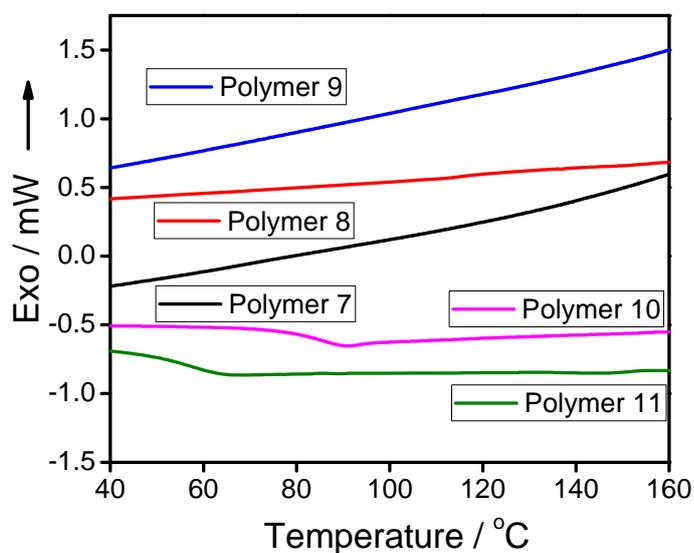

**Figure S3.** DSC curves of **7** (black), **8** (red), **9** (blue) and **10** (pink) and **11** (green). Heating rate: 10 K/min.

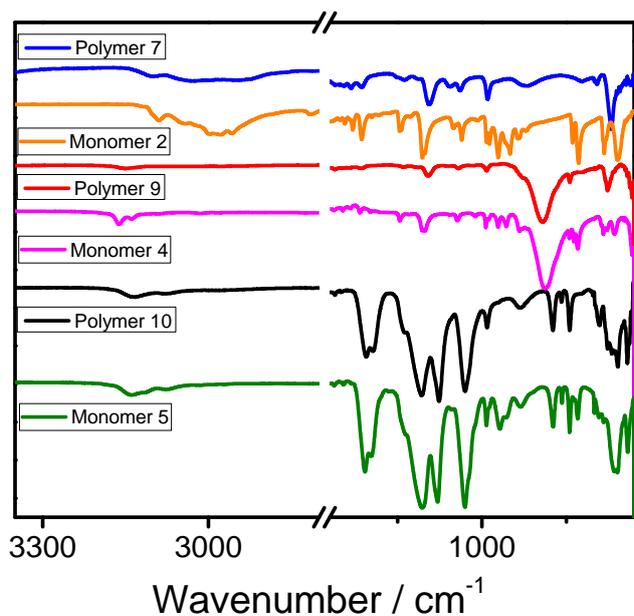

**Figure S4.** FTIR spectra of monomer **2** (orange), **4** (pink) and **5** (green); polymer **7** (blue), **9** (red) and **10** (black).

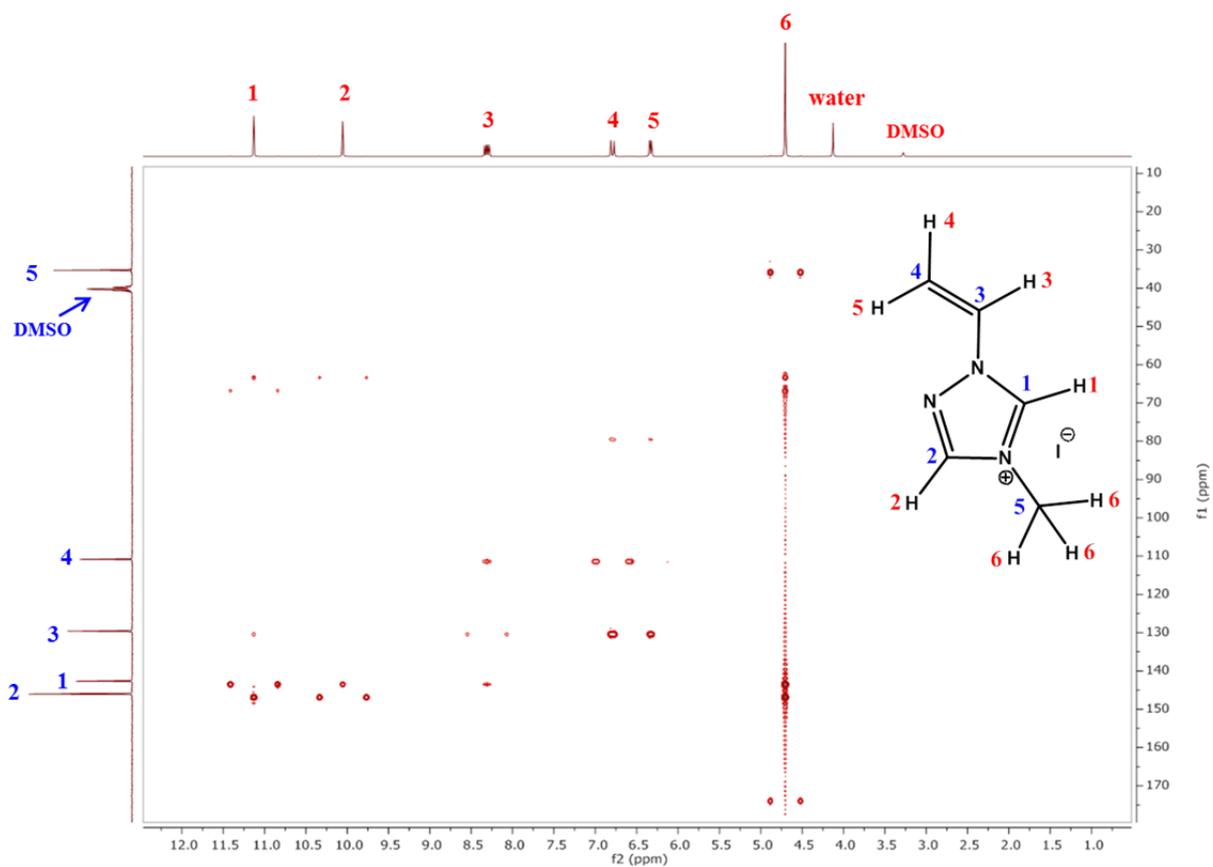

**Figure S5.** $^1$H-$^{13}$C HMBC spectral data of monomer **2**.

*Table S1.* Crystallographic data and structural refinement for monomer **2**

| | | | |
|---|---|---|---|
| Bond precision: | C-C = 0.0180 Å | | Wavelength=0.71073 |
| Cell: | a=5.0409(5) | b=25.591(4) | c=6.9793(8) |
| | alpha=90 | beta=103.817(11) | gamma=90 |
| Temperature: | 293 K | | |
| | Calculated | | Reported |
| Volume | 874.29(19) | | 874.30(19) |
| Space group | P 21/c | | P 1 21/c 1 |
| Hall group | -P 2ybc | | -P 2ybc |
| Moiety formula | C5 H8 N3, I | | C5 H8 N3, I |
| Sum formula | C5 H8 I N3 | | C5 H8 I N3 |
| $M_r$ | 237.04 | | 237.04 |
| $D_x$, g cm$^{-3}$ | 1.801 | | 1.801 |
| Z | 4 | | 4 |
| $M_u$ (mm$^{-1}$) | 3.592 | | 3.592 |
| F000 | 448.0 | | 448.0 |
| F000' | 446.22 | | |
| h,k,l$_{max}$ | 6,30,8 | | 6,30,8 |
| $N_{ref}$ | 1601 | | 1584 |
| $T_{min}$, $T_{max}$ | 0.386,0.650 | | 0.692,1.000 |
| $T_{min}$' | 0.305 | | |
| Correction method= # Reported T Limits: $T_{min}$=0.692 $T_{max}$=1.000 | | | |
| AbsCorr = MULTI-SCAN | | | |
| Data completeness= 0.989 | | Theta(max)= 25.350 | |
| R(reflections)= 0.0374( 1217) | | wR$_2$(reflections)= 0.0808( 1584) | |
| S = 1.062 | | $N_{par}$= 83 | |

*Table S2.* Molecular weight and dispersity of polymer **7** and **8**.

| Polymer | $M_n$ / g/mol | $M_w$ / g/mol | PDI |
|---|---|---|---|
| **7** | 47283 | 89919 | 1.90 |
| **8** | 55558 | 104950 | 1.88 |

*Table S3.* Iodide content and calculated anion exchange degree of monomers and polymers after anion metathesis (calculated from silver nitrate titration with Eosin Y as indicator).

| Compound | Iodide content / wt% | Anion exchange degree / % |
| --- | --- | --- |
| Monomer | | |
| **4** | 0.60 | 98.80 |
| **5** | 0 (no precipitate) | 100 |
| **6** | 0 (no precipitate) | 100 |
| Polymer | | |
| **9** | 2.63 | 92.08 |
| **10** | 1.50 | 95.53 |
| **11** | 1.28 | 96.06 |